\documentclass[a4paper,table,11pt]{article}
\usepackage{jheppub} 

\usepackage{amsmath}
\usepackage{amsfonts}
\usepackage{bm}
\usepackage{mathrsfs}
\usepackage{dsfont}
\usepackage{subfig} 
\usepackage{empheq}
\usepackage{esint}
\usepackage{hyperref}
\usepackage[shortlabels]{enumitem}
\usepackage{tikz, pgf} 
\usetikzlibrary{arrows}
\usepackage{xcolor}
\usepackage{pgfplots}
\usepackage{natbib} 
\usepackage{braket}
\usepackage{slashed}
\usepackage{bbold}

\usepackage[utf8]{inputenc}
\usepackage{array}
\usepackage{ctable}

\newcommand{\fr}[2]{\frac{#1}{#2}}

\newcommand{\del}{\partial}

\newcommand{\w}{\omega}

\newcommand{\Ctor}{\zeta}

\newcommand{\FD}{n_{\mathrm{FD}}}

\newcommand{\ProCtor}{\mathcal{P}^{\Ctor}}

\begin{document}

\title{Holographic KMS relations at finite density}
\author[a]{R. Loganayagam,}
\author[b]{Krishnendu Ray,}
\author[a]{Shivam K. Sharma,}
\author[a]{Akhil Sivakumar.}

\emailAdd{nayagam@icts.res.in}
\emailAdd{krishnendu.ray@physics.ox.ac.uk}
\emailAdd{shivam.sharma@icts.res.in}
\emailAdd{akhil.sivakumar@icts.res.in}

\affiliation[a]{ International Centre for Theoretical Sciences (ICTS-TIFR),
	Tata Institute of Fundamental Research,
	Shivakote, Hesaraghatta,
	Bangalore 560089, India.}
\affiliation[b]{Rudolf Peierls Centre for Theoretical Physics, University of Oxford, \\ Parks Road, Oxford, OX1 3PU, United Kingdom.}

\abstract{We extend the holographic Schwinger--Keldysh prescription introduced in \cite{Glorioso:2018mmw} to charged black branes, with a view towards studying Hawking radiation in these backgrounds. Equivalently we study the real time fluctuations of the dual CFT held at finite temperature and finite chemical potential. We check our prescription using charged Dirac probe fields. We solve the Dirac equation in a boundary derivative expansion extending the results in \cite{Loganayagam:2020eue}. The Schwinger--Keldysh correlators derived using this prescription automatically satisfy the appropriate KMS relations with Fermi--Dirac factors.}

\maketitle

\section{Introduction}\label{sec:intro}
Black holes radiate and their Hawking radiation closely mimic the fluctuations in a thermal bath. Interacting fields in black hole backgrounds can hence be used to model interacting thermal environments. There is however a technical obstacle to such an endeavour: one needs a practical formalism to compute interactions between ingoing quasi--normal modes and the outgoing Hawking fluctuations. Recently, a gravitational Schwinger--Keldysh (grSK) geometry 
\cite{Chakrabarty:2019aeu,Jana:2020vyx} has emerged as an arena where such computations can be performed with ease. This prescription is built on earlier work on real time holography \cite{Son:2002sd, Herzog:2002pc, Son:2009vu, Skenderis:2008dg, Skenderis:2008dh, vanRees:2009rw, Leigh:2009eb, Giecold:2009tt, Barnes:2010ev, Barnes:2010jp, Botta-Cantcheff:2017qir, deBoer:2018qqm,Glorioso:2018mmw}.

The Schwinger--Keldysh, or `in--in', formalism \cite{Schwinger:1960qe,Keldysh:1964ud} is the most robust setting to study the real--time dynamics of non--equilibrium systems \cite{Chou:1984es, kamenev_2011, Bellac:2011kqa, Rammer:2007zz, Landsman:1986uw}. The central idea of the construction involves a path integral with every degree of freedom doubled, describing the evolution of a general non--equilibrium mixed state. For near--equilibrium mixed states,
such a path integral can be computed by a dual gravitational saddle
built by smoothly glueing two copies of the exterior of a black hole across their future horizons. This is the gravitational Schwinger--Keldysh (grSK) geometry alluded to above. While an \emph{ab initio} derivation of this prescription is still not known, its validity and efficacy have been demonstrated in several instances \cite{Glorioso:2018mmw,Chakrabarty:2019aeu,Jana:2020vyx, Loganayagam:2020eue}.

In an accompanying work \cite{Loganayagam:2020eue}, a subset of the authors of this note show that for probe Dirac fermions, the grSK geometry reproduces the correct Fermi--Dirac statistical factors in real time correlations. This setup can be used to compute the  influence phase of an external fermion probing the dual CFT at finite temperature. A natural generalisation of this work is to ask what happens if we consider near--equilibrium states with finite density/chemical potential. Such a generalisation is necessary to make contact with models of holographic condensed matter \cite{Iqbal:2011ae, Hartnoll:2016apf, Hartnoll:2009sz, McGreevy:2009xe, Herzog:2009xv, Doucot:2017bdm, Lee:2009epi, Mross:2010rd, Hartnoll:2007ih, zaanen_liu_sun_schalm_2015}. On the gravitational side, the relevant physics involves the RN--AdS black brane, its quasi--normal modes and its Hawking radiation.  

In this note, we  give a holographic prescription that computes real time correlators at finite chemical potential in the dual CFT. We do this via a geometry made by stitching two copies of 
RN--AdS black brane exteriors which we will term as the RN--SK saddle/geometry. that generalises grSK geometry. In $\S$\ref{sec:grSKq}, we devise a method to write down the outgoing  Hawking  modes in this geometry when the ingoing modes are known. We test this method in $\S$\ref{sec:Diracq} using a probe Dirac field
and show that it indeed yields the correct Fermi--Dirac physics (as well as the correct Kubo--Martin--Schwinger \cite{Kubo:1957mj, Martin:1959jp} relations) at finite chemical potential.
This is followed in $\S$\ref{sec:Deriv} by an explicit solution to Dirac equation upto second order in boundary 
derivative expansion  and the corresponding influence functional. We conclude in  $\S$\ref{sec:conc} with a summary and a discussion of future directions.
  
\section{Hawking radiation in RN AdS backgrounds} \label{sec:grSKq}
In this section, we will describe how to construct Hawking modes in charged black branes when the ingoing quasi--normal modes are known. We will assume that the ingoing solution is naturally specified in ingoing Eddington--Finkelstein coordinates as an analytic function (possibly in  a boundary derivative expansion as is the case, for example, in the fluid--gravity correspondence \cite{Rangamani:2009xk,Hubeny:2011hd}). We will construct these Hawking modes using a bulk $\mathds{Z}_2$ action which is dual to a $CPT$ transformation on the boundary CFT.

Let us begin with the Reissner--Nordstr\"{o}m black brane solution in AdS$_{d+1}$ written out in
ingoing Eddington--Finkelstein time and a \emph{mock tortoise coordinate},  $\zeta $:
\begin{equation}\label{eq:gA}
\begin{split}
&\mathrm{d}s^{2} = - r^{2} \, f \, \mathrm{d}v^{2} + i \, \beta \, r^{2} \, f \, \mathrm{d}v \, \mathrm{d}\zeta + r^{2} \, \mathrm{d}\bm{x}_{d-1}^{2} \, ,\\
& \quad \ \mathcal{A}_{M} \, \mathrm{d}x^{M} = \mathcal{A}_{v} \, \mathrm{d}v = - \mu \left(\frac{r_{h}}{r}\right)^{d-2} \, \mathrm{d}v \, .
\end{split}
\end{equation}
which is the solution of the Einstein-Maxwell bulk action, given by,
\begin{equation}\label{eq:bulkAction}
\frac{1}{2 \kappa^2} \int \mathrm{d}^{d+1}{x} \; \sqrt{-g} \Bigg[\mathcal{R} +d(d-1)-\fr{1}{g_{_{F}}^2} \mathcal{F}_{M N} \mathcal{F}^{M N}\Bigg] \, ,
\end{equation}
where  $g_{_{F}}$ is the gauge coupling constant and we have set the AdS radius to unity. Equivalently, we can replace the description in terms of the metric to one in terms of a set of orthonormal 1--forms \cite{Ceplak:2019ymw, Loganayagam:2020eue},
\begin{equation}\label{eq:tetrads}
\begin{split}
&E^{(v)} = \frac{r}{2}   \, \mathrm{d}v -  \frac{ f}{2} \, r \, \big(i \beta \, \mathrm{d}\zeta - \mathrm{d}v \big), \quad  \ E^{(\zeta )}  = \frac{r}{2}   \, \mathrm{d}v +  \frac{ f}{2} \, r \, \big(i \beta \, \mathrm{d}\zeta - \mathrm{d}v \big), \quad  \ E^{(i)} = r \, \mathrm{d}x^{i} \, .
\end{split}
\end{equation}

In the above equations, $f$ is the emblackening factor of the RN--AdS black brane, $\beta $ is its inverse Hawking  temperature, $r_h$ its outer horizon radius, and $\mu$ its chemical potential. These quantities are, in turn, given in terms of the mass parameter $M$ and the charge parameter $Q$ of the black brane as,
\begin{equation}\label{temp}
\begin{split}
f(r) \equiv 1 +\fr{Q^2}{r^{2d-2}} -\fr{M}{r^d} \, , \qquad
\fr{1}{\beta} \equiv \fr{d\, r_h}{4 \pi}  \left(1- \fr{\left(d-2\right) Q^2}{d \, r_{h}^{2d-2}}\right) \, ,\qquad
\mu  \equiv \frac{g_{_{F}} \, Q}{r_{h}^{d-2}} \left[ \frac{d-1}{2 \left(d-2\right)} \right]^{\frac{1}{2}}\, .
\end{split}
\end{equation}
The ADM mass density $\mathcal{M}$ and the charge density $\mathcal{Q}$ are given by,
\begin{equation}
\mathcal{M} = \frac{(d-1) M}{2 \kappa^2}  \ ,  \qquad  
\mathcal{Q} = \frac{\sqrt{2(d-1)(d-2)}}{g_{_{F}} \, \kappa^2}  Q  \  .
\end{equation} 

\begin{figure}[h]
	\begin{center}
		\begin{tikzpicture}[scale=0.8]

		\draw[white!60!gray, line width = .7] (-4,2.3) -- (-4,-2) ;
		\draw[white!60!gray, line width = .7] (-5,0) --(10,0);
		\draw[cyan!40!teal, ultra thick] (1,1) -- (8,1);
		\draw[cyan!40!teal, ultra thick] (1,1) arc (45: 315 : 1.414);
		\draw[cyan!40!teal, ultra thick] (1,-1) --(8,-1);
		\draw[decoration={snake, segment length = 2.5 mm, amplitude=3}, decorate, red!80!black, thick] (0,0) -- (8,0);
		\draw[decoration={snake, segment length = 2.5 mm, amplitude=3}, decorate, red!80!black, thick] (-3,0) -- (-5,1);

		\node[circle,scale= 0.4,red!80!black,fill] at (-3,0) {} ;
		\node[circle,scale= 0.4,red!80!black,fill] at (0,0) {} ;
		\node[circle,scale= 0.4,red!80!black,fill] at (8,0) {} ;
		\node[circle, fill, cyan!40!teal, scale= 0.17] at (1,1) {} ;
		\node[circle, fill, cyan!40!teal, scale= 0.17] at (1,-1) {};
		\node[circle, fill, black, scale= 0.4] at (8,1) {} ;
		\node[circle, fill, black, scale= 0.4] at (8,-1) {} ;
		
		\node[white!60!gray] at (-4,2.3) {\scriptsize$\blacktriangle$};
		\node[white!60!gray] at (10,0) {\scriptsize $\blacktriangleright$};
		\node[blue!40!black] at (5,1) {$\blacktriangleleft$};
		\node[blue!40!black] at (5,-1) {$\blacktriangleright$};
		\node[blue!40!black] at (-1.41,0) {$\blacktriangledown$};

		\node[teal!40!black] at (4,1.5) {$\mathbf{Re} \,  \zeta = 0$};
		\node[teal!40!black] at (4,-1.5) {$\mathbf{Re} \, \zeta  = 1$};
		\node[red!65!black] at (0,-0.4) {$r_h$};
		\node[red!65!black] at (-3,-0.4) {$r_h^{\mathrm{int}}$};
		\node[red!65!black] at (8,-0.4) {$\infty$};
		\node at (9.7,1) {$\infty+i\varepsilon, \, \Ctor=0$};
		\node at (9.7,-1) {$\infty-i\varepsilon, \, \Ctor=1$};
		\node[white!20!gray] at (-4, 2.75) {\scriptsize $\mathbf{Im} \, r$};
		\node[white!20!gray] at (10.7,0) {\scriptsize $\mathbf{Re} \, r$};
		\end{tikzpicture}
		\caption{\label{fig:fixedv} The radial contour drawn on the complex $r$ plane, at fixed $v$. The locations of the two boundaries and the two horizons have been indicated.}
	\end{center}
\end{figure}
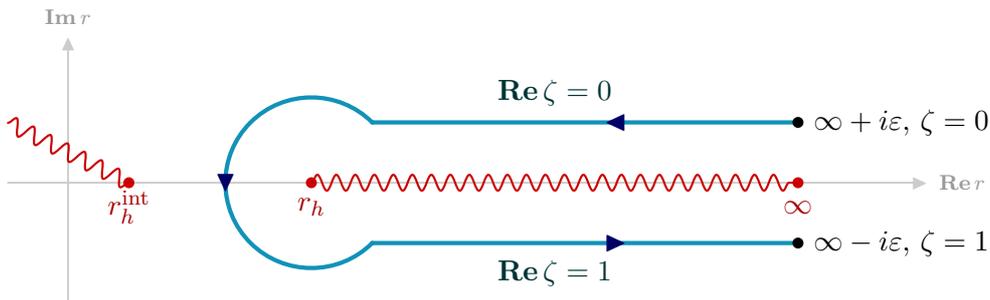

The mock tortoise coordinate $\zeta$ is related to the standard radial coordinate $r$ via the following differential equation,
\begin{equation}\label{mock tortoise}
\begin{split}
\frac{\mathrm{d}r}{\mathrm{d}{\zeta }} = \frac{i \beta }{2} \, r^{2} f(r) \, .
\end{split}
\end{equation}
Since the RHS has simple zeroes at the inner and outer horizons, the coordinate $\zeta$, when thought of as a complex function on the complex $r$ plane, has these
points (along with infinity and other complex simple zeroes of $f(r)$) as branch points. We will
define $\zeta$ by taking a branch cut to extend from the outer horizon to the asymptotic boundary along the real axis. We choose the inner horizon branch cut such that $\zeta$ is analytic in the real interval
between the two horizons (see Fig.\ref{fig:fixedv}). The differential equation above then normalises $\zeta$ such that it has a unit jump across the logarithmic branch cut. Without loss of generality, we will 
take $\zeta (r=\infty + i \varepsilon)=0$. This condition along with the above differential equation then uniquely defines $\zeta$ everywhere in the neighbourhood of $r \in [r_{h}, \infty)$. 

We define the RN--SK geometry as one constructed by taking the RN--AdS exterior and replacing the radial interval
extending from the outer horizon to infinity by a doubled contour, as indicated in Fig.\ref{fig:fixedv}. We then obtain a geometry with two copies of RN--AdS exteriors smoothly stitched together by a `horizon cap' region. This spacetime requires one further identification --- each radially constant slice must meet at the future turning point, $v \rightarrow \infty$.

.

Now that we have elaborated on the RN--SK geometry and its associated notation, let us return to the question posed at the beginning of this section: given an ingoing solution written in ingoing EF coordinates, how do we construct the outgoing Hawking modes? We will do this by exploiting  the $CPT$ symmetry of the boundary CFT which acts as,
\begin{equation}\label{CFT_CPT}
\begin{split}
v \longmapsto - v \, , \ \qquad \bm{x} \longmapsto - \bm{x} \, ,
\end{split}
\end{equation}
in addition to a charge conjugation.  The reader can quickly verify that the transformations as written above do not preserve the RN--AdS solution given in \eqref{eq:gA} and \eqref{eq:tetrads}. So how then  do we implement this symmetry on the RN--AdS solution?

The answer to this conundrum lies in the fact that, by the standard rules of AdS/CFT,  the boundary symmetries can always  be composed with bulk gauge symmetries (since they act trivially on the Hilbert space of states of the CFT). In  this case, we combine the boundary $CPT$ transformation with bulk diffeomorphisms, local Lorentz transformations and gauge transformations. In fact, the 
reader can check that the following transformations preserve the RN--AdS solution: we combine a
charge conjugation $\mathcal{A}_M  \mapsto - \mathcal{A}_M$, a diffeomorphism, 
\begin{equation}\label{AdS_CPTdiffeo}
\begin{split}
v \longmapsto i\beta \zeta - v \, , \ \qquad \bm{x} \longmapsto - \bm{x} \, , 
\end{split}
\end{equation}
a gauge transformation,  
\begin{equation}\label{AdS_CPTgauge}
\begin{split}
\mathbb{\Lambda}\equiv\ -i\beta \int _{0}^{\zeta } \mathrm{d}\zeta' \  \mathcal{A}_{v}(\zeta') \, , 
\end{split}
\end{equation}
and a non--orthochronous local Lorentz transformation,
\begin{equation}\label{AdS_CPT_LLT}
\begin{split}
\mathcal{T}^{a}_{\ \ b} \equiv \begin{pmatrix}
-1 & 0 & 0 \\
0 & 1 & 0 \\
0 & 0 & -\mathds{1}
\end{pmatrix}\begin{pmatrix}
\cosh \vartheta  & \ \sinh \vartheta  & \ 0
\\
\sinh \vartheta  & \ \cosh \vartheta  & \ 0
\\
0 & \ 0 & \ \mathds{1}
\end{pmatrix}, \ \ \vartheta \equiv \log f.
\end{split}
\end{equation}
The linear transformation $\mathcal{T}$ is an idempotent local Lorentz transformation (LLT) composed of a boost along $\zeta$ with rapidity $\vartheta$ followed by a reflection of the $v$ and $\bm{x}$ axes.

To check the invariance of the geometry, we will write the  diffeomorphism in Eq.\eqref{AdS_CPTdiffeo} in the form $X^A\mapsto  X^B \, \mathcal{J}^{\ \, A}_{B}$ where,
\begin{equation}
\begin{split}
\mathcal{J}^{\ \, B}_{A}\equiv \begin{pmatrix}
-1 & 0 & 0 \\
i \beta & 1  & 0 \\
0 & 0 & -\mathds{1}
\end{pmatrix}.
\end{split}
\end{equation}
The invariance is then the statement that 
\begin{equation}
\begin{split}
\mathcal{T}^{a}_{\ \ b}  \; \mathcal{J}^{\ \, B}_{A} \; E^{b}_{B}   = E^{a}_{A}\ ,\qquad \qquad
-\left(\mathcal{J}^{\ \, M}_{N} \; \mathcal{A}_{M} +\partial_N \mathbb{\Lambda}\right)  = \mathcal{A}_{N}\ ,
\end{split}
\end{equation}
as can be readily checked. 

It is instructive to unify these transformations by  imagining the gauge field as descending from a Kaluza--Klein reduction over a circle. Say there was an extra circle
direction denoted by $\varphi$, along which we have a 1--form, $\mathrm{d}\varphi+\mathcal{A}_N \, \mathrm{d}X^N$, that is orthonormal to the set of 1--forms, $E^{a}_{M} \, \mathrm{d}X^M$. The action of  charge conjugation 
and gauge transformation in this picture can then be thought of as the KK diffeomorphism $\varphi \mapsto \mathbb{\Lambda}-\varphi$ (which extends $\mathcal{J}^{\ \, B}_{A}$ to the  $\varphi$ direction)  along with a KK LLT that acts as a reflection along the $\varphi$ axis (which extends $\mathcal{T}^{a}_{\ \ b} $ to the $\varphi$ direction). The mathematically inclined 
reader would recognise that we have essentially given a description of the boundary $CPT$ transformation as an
automorphism of the RN--AdS principle bundle.

Having implemented the boundary $CPT$ in the bulk, we will now use  this symmetry to generate new solutions to the  bulk field equations linearised about the RN--AdS background. To see how this is done, let us begin with the simplest example of a complex scalar field with  charge $q$ probing the RN--AdS black brane. In the Fourier domain, we can expand a field configuration  in terms of planar waves:
\begin{equation}
\begin{split}
\int \frac{\mathrm{d}{\omega } \ \mathrm{d}^{d-1}{\bm{k}}}{(2 \pi )^{d}} \ \Phi (q,\omega , \zeta , \bm{k}) \, e^{- i \omega v + i \bm{k} \cdot \bm{x}} \, .
\end{split}
\end{equation}
Say this configuration solves the diffeomorphism/gauge covariant linear PDE in the bulk. How do we use the $\mathds{Z}_2$ we constructed above to get new solutions? The quickest way to answer this is to use the KK (or the principal bundle) picture described above. To 
this end, we think of the above field as drawn from a KK decomposition of a higher dimensional scalar field of the form, 
\begin{equation}
\begin{split}
\sum_{q\in\mathds{Z}} e^{iq\varphi} \int \frac{\mathrm{d}{\omega } \ \mathrm{d}^{d-1}{\bm{k}}}{(2 \pi )^{d}} \ \Phi (q,\omega , \zeta , \bm{k}) \, e^{- i \omega v + i \bm{k} \cdot \bm{x}} \, .
\end{split}
\end{equation}
The diffeomorphism/gauge covariant linear PDE can then be lifted to a PDE covariant under higher dimensional diffeomorphisms. We perform a higher dimensional diffeomorphism $\{v \mapsto i\beta \zeta - v\ ,\bm{x} \mapsto - \bm{x}\ ,\varphi \mapsto \mathbb{\Lambda}-\varphi\}$ on the Fourier domain expression given above. After relabelling the sum/integrals, we obtain,
\begin{equation}
\begin{split}
\sum_{q\in\mathds{Z}} e^{iq\varphi} \int \frac{\mathrm{d}{\omega } \ \mathrm{d}^{d-1}{\bm{k}}}{(2 \pi )^{d}} \  e^{-\beta \omega \zeta -iq\mathbb{\Lambda} }\ \Phi(-q,-\omega , \zeta , -\bm{k}) \, e^{- i \omega v + i \bm{k} \cdot \bm{x}} \, .
\end{split}
\end{equation}
The higher dimensional diffeomorphism covariance of the linear PDE then guarantees that the map
\begin{equation}
\Phi(q, \omega , \zeta, \bm{k}) \longmapsto e^{-\beta \omega \zeta -iq\mathbb{\Lambda} }\ \Phi (-q,-\omega , \zeta , -\bm{k}) 
\end{equation} 
generates new solutions.  In fact, the pre--factor in the exponent shows that this map takes
ingoing solutions that are analytic in $r$ to outgoing solutions which exhibit branch cuts.
We can thus generate the Hawking modes from quasi--normal modes via the map,
\begin{equation}
\Phi^{\mathrm{in}}(q, \omega , \zeta, \bm{k}) \longmapsto e^{-\beta \omega \zeta  -iq\mathbb{\Lambda} }\  \Phi^{\mathrm{in}}(-q,-\omega , \zeta , -\bm{k}) \equiv \Phi ^{\mathrm{out}}(q, \omega ,\zeta ,\bm{k})\ .
\end{equation} 

The extension to other fields is straightforward. As a pertinent example, let us consider how this works for charged Dirac spinors. The main novelty in this case is the additional action of LLTs on spinor indices. The complex boost given in \eqref{AdS_CPT_LLT} acts on spinor indices via 
\begin{equation}
\begin{split}
\mathfrak{T} \equiv \Gamma ^{(\zeta )} \cdot \exp\Big(\, \frac{\vartheta }{2} \ \Gamma ^{(\zeta )} \, \Gamma ^{(v)}\Big) = \sqrt{f} \ \mathbb{\Gamma } + \frac{1}{\sqrt{f}} \, \mathbb{\Gamma }^{\dagger}\, ,
\end{split}
\end{equation}
where we have defined,
\begin{equation}
\begin{split}
\mathbb{\Gamma } \equiv \frac{1}{2} \left(\Gamma ^{(v)} + \Gamma ^{(\zeta )}\right).
\end{split}
\end{equation}
The Gamma matrix notation here is standard and our convention for the Clifford algebra is $\left\{\Gamma ^{a}, \Gamma ^{b}\right\} = 2 \, \eta ^{ab} \, \mathds{1}$, with a mostly positive signature
$\eta ^{ab}$. The map from ingoing to outgoing solutions, in this case, is given by,
\begin{equation}\label{eq:spinorCPT}
\begin{split}
\Psi ^{\mathrm{in}}(q, \omega ,\zeta ,\bm{k}) \longmapsto e^{-\beta \omega \zeta -iq\mathbb{\Lambda} }\ \mathfrak{T} \cdot \Psi ^{\mathrm{in}} (-q, - \omega ,  \zeta ,- \bm{k}) \, \equiv \Psi ^{\mathrm{out}}(q, \omega  ,\zeta , \bm{k}) \, .
\end{split}
\end{equation}

In order to streamline the analysis that is to follow, we end this section with a discussion of how the quantities that we have introduced behave as they traverse the horizon cap. As noted before, the mock tortoise coordinate has a unit jump when it encircles $r = r_{h}$. Noting that the gauge field $\mathcal{A}_{v}$ has the value $-\mu $ at the horizon, we see that the jump in $e^{-iq\mathbb{\Lambda} }$ is the  fugacity of the boundary theory,
\begin{equation}
\begin{split}
\lim_{\zeta \rightarrow 0} e^{-iq\mathbb{\Lambda} } = 1, \qquad \quad \lim_{\zeta \rightarrow 1} e^{-iq\mathbb{\Lambda} } = e^{q\beta \mu } \equiv z \, .
\end{split}
\end{equation} 
The rapidity parameter $\vartheta$, on the other hand, has a jump of $2 \pi i$ when it encircles the outer horizon --- this can be traced to the phase $e^{2 \pi i}$ that the function $f$ picks up around the same path. As a result, every spinor index receives an additional `fermionic'  minus sign as it crosses the horizon cap. For a field of spin $s$,
one gains the statistical factor $(-1)^{2s}e^{-\beta\omega+q\beta\mu}$ which is identical 
to the factor gained in Euclidean path integrals as we traverse the thermal circle. As we will argue below, this factor is crucial to how the RN--SK geometry gets the physics of Hawking radiation right.

\section{The Dirac equation}\label{sec:Diracq}
We will now apply the preceding slightly abstract discussion to the explicit example of a 
Dirac equation in the RN--SK background. The action for a bulk Dirac spinor $\Psi $ of charge $q$ and mass $m$ living on the RN--SK geometry is given by,
\begin{equation}\label{bulk action}
\mathds{S}_{\Psi } = \ointctrclockwise \mathrm{d}\Ctor \, \int \mathrm{d}^{d}x \; \sqrt{-g} \ i \, \overline{\Psi}  \left( \Gamma^{M} \, \mathbb{D}_{M} - m \right) \Psi +  \int \mathrm{d}^{d}x \;  i \,  \overline{\Psi} \, \ProCtor_{-}  \, \Psi \ \Bigg|^{\Ctor=1}_{\Ctor=0} \ .
\end{equation}

Let us explain each of the above terms. The first is the standard bulk Dirac action, that fixes the equations of motion to be the Dirac equation in this curved spacetime. We have introduced above the covariant Dirac derivative,
\begin{equation}\label{covariant Dirac derivative}
\mathbb{D}_{M} \equiv \del_{M} - i q  \mathcal{A}_{M} + \fr{1}{4} \w_{ab \, M}\Gamma^{a}\Gamma^{b} \ ,
\end{equation}
where $\mathcal{A}_{M}$ is the background gauge potential and $\w_{ab \, M}$ are the spin--connection 1--forms derived from Eq.\eqref{eq:tetrads}. 

The second term is a variational boundary term, added in to ensure that this action admits a well--posed variational principle when the spinors satisfy semi--Dirichlet boundary conditions \cite{Henningson:1998cd, Mueck:1998iz, Henneaux:1998ch, Iqbal:2009fd} (more will be said about this below). When the above action is evaluated on--shell, the bulk term vanishes and only the boundary term contributes to the final answer. Notice that the boundary term receives contributions from both the left and right boundaries of the RN--SK saddle \cite{Loganayagam:2020eue}. The operator $\mathcal{P}^{\zeta }_{-}$ involved in the definition of this term is one of two projection operators, defined as,
\begin{equation}
\ProCtor_{\pm} \equiv \fr{1}{2}\left(\mathds{1}\pm\Gamma^{(\Ctor)} \right) \ .
\end{equation}

The Dirac equation $\left(\Gamma^{M} \, \mathbb{D}_{M}-m\right)\Psi = 0$, written out in the Fourier domain, takes the following form:
\begin{equation}\label{eq:DiracEquation}
\left\{\mathbb{\Gamma} \left(\del_{\Ctor} + \beta \w + \beta q \mathcal{A}_{v}+ \fr{1}{2}\del_{\Ctor} \ln f  \right) +\mathbb{\Gamma}^{\dagger}f^{-1}\del_{\Ctor} - \fr{ \beta}{2}\left(\Gamma^{(i)}k_{i}+imr \right) \right\} \left(r^{\fr{d}{2}} \Psi \right) = 0 \ .
\end{equation}
It can be explicitly checked that if $\Psi (q, \omega , \zeta , \bm{k})$ solves this equation, applying the transformation described in \eqref{eq:spinorCPT} to it generates another solution 
of the same equation.

The relevant AdS/CFT boundary conditions  are,  
\begin{equation}\label{eq:SK-GKPW}
\lim_{\Ctor \to 0} r^{\fr{d}{2}-m} \, \ProCtor_{+} \, \Psi = \ProCtor_{+}S_{0} \, \psi_L  \, ,\qquad \quad \lim_{\Ctor \to 1} r^{\fr{d}{2}-m} \,  \ProCtor_{+} \,  \Psi = \ProCtor_{+}S_{0} \, \psi_R \, .
\end{equation}  
Here $S_{0}$ is a constant boundary--to--bulk matrix \cite{Loganayagam:2020eue}. It is defined in terms of the ingoing boundary-to-bulk propagator $S^{\text{in}}(\w,\Ctor,\bm{k})$ as,
\begin{equation}\label{eq:DefnS0}
S_{0} \equiv \lim_{\Ctor \to 0} r^{\fr{d}{2}-m} \ S^{\text{in}}(\w,\Ctor,\bm{k}) =\lim_{\Ctor \to 1} r^{\fr{d}{2}-m} \ S^{\text{in}}(\w,\Ctor,\bm{k}) \ .
\end{equation}
The above boundary conditions are a `doubled' version of the standard semi--Dirichlet conditions in AdS/CFT. These boundary conditions uniquely determine the 
solution on the RN--SK geometry.

To see this, we begin with the most general combination of ingoing and outgoing solutions:
\begin{equation}
\begin{split}
\Psi (q,\omega ,\zeta ,\bm{k}) &= - S^{\mathrm{in}}(q,\omega ,\zeta ,\bm{k}) \, \psi _{\bar{F}}(q, \omega , \bm{k})- S^{\mathrm{out}}(q,\omega ,\zeta ,\bm{k}) \, \psi _{\bar{P}}(q, \omega , \bm{k}) \, e^{\beta \left(\omega  - q\mu \right)} \ ,\\
&= - S^{\mathrm{in}}(q,\omega ,\zeta ,\bm{k}) \, \psi _{\bar{F}}(q, \omega , \bm{k})- S^{\mathrm{rev}}(q,\omega ,\zeta ,\bm{k}) \, \psi _{\bar{P}}(q, \omega , \bm{k}) \, e^{\beta \omega(1-\zeta)+ iq\left(i\beta \mu - \mathbb{\Lambda}\right)} \, .
\end{split}
\end{equation}
Here, we have defined,  
\begin{equation}
\begin{split}
S^{\mathrm{rev}}(q,\omega ,\zeta ,\bm{k})\equiv S^{\mathrm{in}}(-q,-\omega ,\zeta ,-\bm{k})\, , 
\end{split}
\end{equation}
and we have used the $\mathds{Z}_2$ action detailed in the last section to get the outgoing Hawking solution.

The boundary conditions described above fix  
\begin{equation}
\begin{split}
\psi _{\bar{F}} \equiv \FD \left(\psi _{R} - \psi _{L}\right) - \psi _{R}\, , \qquad \quad \psi _{\bar{P}} \equiv \FD \left(\psi _{R} - \psi _{L}\right) \ 
\end{split}
\end{equation}
with
\begin{equation}
\FD \equiv \fr{1}{e^{\beta (\w-q\mu)}+1} \, 
\end{equation}
being the familiar Fermi--Dirac factor at finite chemical potential. We recognise here 
the \emph{retarded--advanced basis} (RA) for the boundary spinor sources \cite{Chou:1984es,
	Chaudhuri:2018ymp}. Thus, when written in the RA basis, the two combinations of sources precisely multiply the ingoing/quasi–normal bulk–to–boundary propagator and the outgoing bulk–to–boundary propagator respectively. This is analogous to the corresponding statements
at zero chemical potential \cite{Son:2009vu,Chakrabarty:2019aeu,Jana:2020vyx,Loganayagam:2020eue}. 

As pointed out in \cite{Loganayagam:2020eue}, the above fact can be used to argue why the generating function of correlations computed using our holographic prescription satisfy both the Schwinger--Keldysh \emph{collapse rules} as well as the KMS conditions, i.e., the generating function cannot contain terms with only $\psi _{\bar{P}}$ or terms with only $\psi _{\bar{F}}$. Exactly the same argument is valid for the case of finite chemical potential presented here. We direct the reader to \cite{Loganayagam:2020eue} for more details.

Finally, in the \emph{Keldysh--rotated} or \emph{average--difference} basis, one gets,
\begin{equation}
\begin{split}
\Psi (v, \zeta , \bm{k}) = S^{\mathrm{in}} \,\psi _{a} -\left\{ \left( \FD-\frac{1}{2}\right)  S^{\mathrm{in}} + \FD \ e^{\beta \omega(1-\zeta)+ iq \left(i\beta \mu - \mathbb{\Lambda}\right)} \ \mathfrak{T} \cdot S^{\mathrm{rev}} \,  \right\} \psi _{d} \, .
\end{split}
\end{equation}
This basis makes the Schwinger--Keldysh collapse rules manifest, i.e., there are no terms in the generating function for boundary correlators that have only $\psi _{a}$ terms. Equivalently, correlation functions composed of only the \emph{difference} operator $\mathcal{O}_{d}$ vanish. As we will see in $\S$\ref{sec:Deriv}, the influence phase that we derive for the open EFT precisely has this property.

\section{Gradient expansion}\label{sec:Deriv}
In this section, we will find the solution to the Dirac equation in gradient expansion, generalizing the work of \cite{Loganayagam:2020eue} to \emph{charged} black branes. We only consider the case of a massless spinor field for simplicity.

We begin by writing the most general solution $\Psi \equiv \Psi (v ,\zeta, \bm{x})$ in a gradient expansion, compatible with rotational invariance as,
\begin{equation}\label{Massless Ansatz}
\begin{split}
\Psi  = \frac{ 1 }{ r^{d/2} }  \Big\{ \mathds{1} + C_{a}^{(1)} M^a  \del_v +  D_{a}^{(1)} M^a \Gamma^{(i)} \del_i + C_{a}^{(2)} M^a \del_v^{2} + D_{a}^{(2)} M^a \del_i^{2} + \dots \Big\} S_0  \,  \psi
\end{split}
\end{equation}
where $\psi \equiv \psi(v,\bm{x})$ is the boundary spinor acting as a source at the boundary. The functions $\{ C^{(\mathrm{i})}, D^{(\mathrm{j})},  \dots  \}$ are some unknown functions of the radial direction $\zeta$. The matrices $M^{a}$ that appear in the above equation belong to the set $ \{\mathds{1},\Gamma^{(\zeta)}  \}$ and in order to satisfy the Dirac equation \eqref{eq:DiracEquation} at zero derivative order, we will fix $S_0$ to be a constant matrix annihilated by \(\mathbb{\Gamma }\), i.e., $\mathbb{\Gamma }\,S_0 = 0$.

The bulk--to--boundary Green's function satisfying ingoing boundary conditions can be computed by 
substituting the above ansatz into the Dirac equation and solving it order by order in 
derivative expansion. In the Fourier domain we get an expression of the form,
\begin{equation}\label{Inansatz}
\begin{split}
S^{\mathrm{in}}  = \frac{1}{r^{d/2}}\Bigg\{ 1 + \frac{\beta}{2}&\Big( \; H\, \Gamma^{(\zeta)} - H_{(0)} \Big)\Gamma^{(i)}\,k_i\, -\frac{\beta^2 \,\omega}{2}\Big( \widetilde{H}\;\Gamma^{(\zeta)}- \widetilde{H}_{(0)} \Big)\Gamma^{(i)}\,k_i\, \\
-&\frac{\beta^2\, \bm{k}^2}{8}\,  \Big( f \,  G^2 - f_{(0)}\, G_{(0)}^{2}\Big) \, -\frac{\beta^2\, \bm{k}^2\, H_c}{4}\Big( \; H\, \Gamma^{(\zeta)} - H_{(0)} \Big)  +\ldots\Bigg\}S_0 \ .
\end{split}
\end{equation}
where \(\bm{k}^2 =k_i k^i\) and the subscript `\((0)\)' denotes the value of the functions at the conformal boundary, $\zeta=0$. The unknown functions \(H\), \(\widetilde{H}\), and \(G\) must satisfy the following radial differential equations:
\begin{equation}\label{unknownfunctions}
\begin{split}
\frac{d}{d\zeta}\left( e^{iq\mathbb{\Lambda}} \, \sqrt{f} \, H \right) & = e^{iq\mathbb{\Lambda}} \, \sqrt{f} \ ,\quad
\frac{d}{d\zeta}\left( e^{iq\mathbb{\Lambda}} \, \sqrt{f} \, \widetilde{H} \right)  = e^{iq\mathbb{\Lambda}} \, \sqrt{f}\ H \ ,\quad
\frac{d}{d\zeta}\left(  \sqrt{f} \, G \right)  =\sqrt{f} \  .
\end{split}
\end{equation}
Since we want an ingoing solution, we take  \(H\), \(\widetilde{H}\), and \(G\) to be analytic at the outer horizon which then implies that the functions
$\sqrt{f} H, \sqrt{f} \widetilde{H}$ and $\sqrt{f} G$ should vanish at the outer horizon.
With this boundary condition specified, the above ODEs have a unique solution. 

Given the above ingoing solution, the corresponding solution on the RN--SK geometry follows
by using the formulae derived in $\S$\ref{sec:Diracq}. We will end this section by quoting the 
result for the influence phase in \emph{average--difference} basis:
\begin{equation} \label{Influence Phase}
\mathds{S}_{\mathrm{IF}}[\psi_a,\psi_d] = \int \mathrm{d}^{d}{x} \ r^{d} \ i \,  \overline{\Psi } \, \mathcal{P}^{\zeta}_{-} \Psi \  \Bigg|^{\zeta = 1}_{\zeta = 0} = \int \mathrm{d}^{d}{x} \,i \Bigg\{\overline{\psi}_a \, \mathcal{S}^{d a}\, \psi_d
+\overline{ \psi}_d\, \mathcal{S}^{a d}\, \psi_a
+\overline{ \psi}_d \, \mathcal{S}^{a a}\, \psi_d \Bigg\}
\end{equation}
where $\mathcal{S}^{ad}$, $\mathcal{S}^{da}$, and $\mathcal{S}^{aa}$ are the retarded, advanced and Keldysh  Green's functions respectively. When the dimension of the boundary theory, $d$, is odd, their explicit forms  are given by,
\begin{equation}
\begin{split}
\mathcal{S}^{ad} &=  \gamma^{(v)}-\beta H_{(0)}\,\gamma^{(i)} k_i+\beta^2  \omega\,\gamma^{(i)} k_i \widetilde{H}_{(0)} +\frac{\beta^2  \bm{k}^2 (H_{(0)})^2}{2}\gamma^{(v)}  + \ldots   \\
\mathcal{S}^{da} &= - \Bigg(\gamma^{(v)}+\beta H_{(0)}^{\star}\,\gamma^{(i)} k_i +\beta^2  \omega\,\gamma^{(i)} k_i \widetilde{H}_{(0)}^{\star} +\frac{\beta^2  \bm{k}^2 (H_{(0)}^{\star})^2}{2}\gamma^{(v)}+ \ldots \Bigg) \\
\mathcal{S}^{aa} &=  e \,\gamma^{(v)}+ \fr{\beta \omega \,(1-e^2)}{2} \, \gamma^{(v)} -\frac{\beta^2 \omega^2\,\big[ e \,(1-e^2) \big]}{4}\gamma^{(v)}+\frac{e\, \beta \,(H_{(0)}^{\star}-H_{(0)} )}{2} \,\gamma^{(i)} k_i \\
& +\fr{\beta^2 \omega}{4} \gamma^{(i)} \, k_i \Bigg(  (1-e^2)\big( H_{(0)}^{\star}-H_{(0)}\big)+2\,e \,(\widetilde{H}^{\star}_{(0)} +\widetilde{H}_{(0)})\Bigg) +\frac{e \, \beta^2 \bm{k}^2\, \,\big[(H_{(0)}^{\star})^2+ H_{(0)}^2\big]}{4} \, \gamma^{(v)}  +\ldots 
\end{split}
\end{equation}
where we have expanded the Fermi--Dirac factors in powers of \(\omega\) and the superscript `$\star$' denotes the replacement $q \rightarrow -q$ in the arguments of the functions. We use the letter $e$ to denote the ground state charge 
\begin{equation} 
e \equiv \frac{1}{1+e^{-q\beta\mu}}- \frac{1}{1+e^{q\beta\mu}}\ ,
\end{equation}
which is the difference between the ground--state occupation of fermionic particles and anti--particles. A similar set of expressions can be obtained $d$ is even, as described in \cite{Loganayagam:2020eue}. This influence phase obeys the 2--point KMS relation which takes the following form in the a-d basis,
\begin{equation}\label{eq:KMSrelation}
\mathcal{S}^{aa} =\fr{1}{2}  \frac{e^{\beta \omega}-z}{e^{\beta \omega}+z}\,\left(\mathcal{S}^{ad}-\mathcal{S}^{da} \right) . 
\end{equation}
In the limit $\omega, \bm{k} \rightarrow 0$, the influence phase has a very simple form,
\begin{equation}
\mathds{S}_{\mathrm{IF}} =\int \mathrm{d}^{d}{x} \  i \Big( \psi_{a}^{\dagger} \psi_d-\psi_{d}^{\dagger} \psi_a -e\,\psi_{d}^{\dagger} \psi_d \Big).
\end{equation}
Finally, in the limit of uncharged black branes or  vanishing chemical potential ($q, \, \mu \rightarrow 0$), all these Green's functions and the influence phase match with the results derived in \cite{Loganayagam:2020eue}.

\section{Conclusions and future directions } \label{sec:conc}

In this note we extended the prescription for gravitational Schwinger--Keldysh saddles to the case of charged black branes. This geometry is dual to the real--time evolution of a CFT held at a finite temperature and chemical potential. Our prescription is a straightforward generalisation of the original grSK prescription \cite{Glorioso:2018mmw, Loganayagam:2020eue, Chakrabarty:2019aeu, Jana:2020vyx}. In this geometry, we describe how to obtain the outgoing or Hawking modes from the ingoing quasi--normal modes via a $CPT$ transformation.

In order to test the above prescription, we probe the RN--AdS black brane with Dirac fermions charged under the gauge field.  Using this solution, we derive the influence phase of a probe fermion interacting with the holographic CFT  \cite{Loganayagam:2020eue}. We show that the correlators  derived using this holographic prescription satisfy the desired KMS relations dressed with the correct Fermi--Dirac factors. This is our central result which generalises the results of \cite{Loganayagam:2020eue} to include the chemical potential of the CFT. Further, specialising to the case of massless fields, we derive the  quasi--normal modes up to second order in a boundary derivative expansion.

A natural follow up to this study would be to examine the near--extremal limit. In this limit, the outer and the inner horizons come close to each other and can lead to the pinching of the holographic radial contour. It would be interesting to see how this should be dealt with and make contact with the derivative expansion in near--extremal branes recently claimed in \cite{Moitra:2020dal}.

\bigskip
\section*{Acknowledgements.}
It is with great pleasure that we thank Bidisha Chakraborty, Greg Henderson, Chris Herzog, Aswin P. M.,  Suvrat Raju, Mukund Rangamani, Omkar Shetye and Spenta Wadia for useful discussions. KR is supported by the Rhodes Trust via a Rhodes Scholarship. The authors would like to acknowledge their debt to the people of India for their sustained and generous support to research in the basic sciences.

\bibliographystyle{JHEP}
\bibliography{RNSKDraft}

\end{document}